\documentstyle[aps,epsf]{revtex} 

\begin{document}

\title{ Decoherence and long-lived Schr\"odinger cats in BEC}
\author{ Diego A. R. Dalvit$^1$
	   \thanks{e-mail: \tt dalvit@lanl.gov}, 
         Jacek Dziarmaga$^{1,2}$
         \thanks{ e-mail: \tt dziarmaga@t6-serv.lanl.gov }, 
         and 
         Wojciech H. Zurek$^1$ \thanks{e-mail: \tt whz@lanl.gov}}
\address{ 1) Los Alamos National Laboratory, T-6, Theoretical Division, 
             MS-B288, Los Alamos, NM 87545, USA                              \\
          2) Institute of Physics, Jagiellonian University, Krak\'ow, Poland }

\date{\today}
\maketitle
\tighten

\begin{abstract} 
{\bf We consider quantum superposition states in Bose-Einstein condensates. 
A decoherence rate for the Schr\"odinger cat is calculated and 
shown to be a significant threat to this macroscopic quantum superposition 
of BEC's.
An experimental scenario is outlined where the decoherence rate due to the
thermal cloud is dramatically reduced thanks to trap engineering and
``symmetrization" of the environment. We show that under the proposed scenario
the Schr\"odinger cat belongs to an approximate decoherence-free pointer
subspace.}

\end{abstract}

\section{Introduction}

Microscopic quantum superpositions are an everyday physicist's
experience.  Macroscopic quantum superpositions, despite nearly a century
of experimentation with quantum mechanics, are still encountered only very
rarely. Fast decoherence of macroscopically distinct states is to be
blamed \cite{zurek}. In spite of that, recent years were a witness to an
interesting quantum optics experiment \cite{paris} on decoherence of 
a few photon
superpositions. Moreover, matter-wave interference in fullerene
$\rm{C}_{60}$ has been observed \cite{insbruck}. Another 
experiment has succeeded in ``engineering'' the environment in the context
of trapped ions \cite{wineland}. Recently the first detection of a 
macroscopic Schr\"odinger cat state in an rf-SQUID was reported
\cite{squid}. All these successes tempt one to push similar investigations
of basic quantum mechanics into the rapidly progressing field of
Bose-Einstein condensation (BEC) of alkali metal atomic vapors \cite{bec}
The condensates can contain up to $10^7$ atoms in the same quantum
state. What is more, it is possible to prepare condensates in two
different internal states of the atoms. Some of these pairs of internal
levels are immiscible, and their condensates tend to phase separate into
distinct domains with definite internal states \cite{spinor}. The
immiscibility seems to be a prerequisite to prepare a quantum
superposition in which all atoms are in one or 
the other internal state, 
$|\psi\rangle = (|N,0\rangle + |0,N\rangle )/\sqrt{2}$, 
where $N$ is the total number of condensed atoms. There are at least
two theoretical proposals how to prepare a macroscopic quantum
superposition in this framework \cite{lz,gs}. 
Other proposals involve non-unitary evolution of the BEC towards the
Schr\"odinger cat state by means of continuous quantum measurements
\cite{r1,r2}. Neither of these proposals addresses
the crucial question of decoherence. 

This paper is a simplified and more pedagogical version of our previous
work \cite{us}. All technical details of the calculations, in particular
the computation of the decoherence rate for the BEC superposition state,
can be found in that reference.

\section{The condensate}

Let us first make a short summary of some of the ideas for creating quantum
superposition states of Bose-Eintein condensates. The methods that have
been thus far proposed start with two weakly interacting dilute Bose 
condensates  of atoms in different internal states. 
Let us call these internal states
$A$ and $B$. Atoms interact through s-wave collisions, characterized by a 
single parameter $a$ which is known as the scattering length. We shall
consider the case of repulsive interaction $a>0$; it is known that
for the opposite case the condensate is unstable above a critical total 
number of particles. It is assumed
that the inter-scattering lengths for collisions $A-A$ and $B-B$ are the same,
i.e. $a_{AA}=a_{BB}$, which is in general different from the intra-scattering
length $a_{AB}$. When the self-energy of atom-atom interactions of the BEC is
much less than the mode energy spacing, then one can treat the condensate in 
the two-mode approximation.
Moreover, the condensate is shined with a appropriately
chosen laser that introduces a Josephson-like coupling of strength $\lambda$ 
that interchanges internal atomic states in a coherent manner. 
The condensate two-mode Hamiltonian is

\begin{equation}\label{HC}
H_{\rm C}=\epsilon_g(a^{\dagger}a+b^{\dagger}b)-
    \lambda(a^{\dagger}b+b^{\dagger}a)+               
    \frac{u_{\rm c}}{2}(a^{\dagger}a^{\dagger}aa+b^{\dagger}b^{\dagger}bb)+
    v_{\rm c}(a^{\dagger}b^{\dagger}ab)\;,
\end{equation}
Here $\epsilon_g$ is the energy of the lowest single-particle state in the
trap (assumed to be the same for $A$ and $B$), 
$u_{\rm c}= 4 \pi \hbar^2 a_{AA}/m$ and $v_{\rm c}= 4 \pi \hbar^2 a_{AB}/m$.

This hamiltonian was studied in great detail in \cite{lz,gs,steel}.
The precise quantum state of the condensate for all values of $u_{\rm c}$
and $v_{\rm c}$ was studied by means of a Wigner-like distribution on the 
two-mode Bloch sphere.
It was found that 
when the repulsion between atoms in different states is bigger
that the repulsion between atoms in the same state, that is when
$v_{\rm c} > u_{\rm c}$, then the condensate tends to phase separate in space
into distinct domains with definite internal states. This condition is known
as the immiscibility condition. More importantly, the ground state of the
system is a quantum superposition state. When the ``purity'' factor defined
as $\epsilon \equiv (\lambda/(v_{\rm c}-u_{\rm c})N)^N $ is much less
than one, then  the lowest
energy subspace contains two macroscopic superpositions

\begin{eqnarray}\label{pure}
&& |+\rangle=
   \frac{1}{\sqrt{2\;N!}}[ (a^{\dagger})^N + (b^{\dagger})^N ] \; |0,0\rangle
   \equiv
   \frac{1}{\sqrt{2}}( |N,0\rangle + |0,N\rangle ) \;\;,\nonumber\\
&& |-\rangle=
   \frac{1}{\sqrt{2\;N!}}[ (a^{\dagger})^N - (b^{\dagger})^N ] \; |0,0\rangle
   \equiv
   \frac{1}{\sqrt{2}}( |N,0\rangle - |0,N\rangle ) \;\;.
\end{eqnarray}
The lower $|+\rangle$ and the higher $|-\rangle$ states are separated by a
small energy gap of $E_- - E_+ = N (u_{\rm c}-v_{\rm c})\epsilon\ln\epsilon$. 
In the extreme case of $\lambda=0$ (no laser applied) the two states are 
completely degenerate. If we just have $\epsilon<1$,
the $|\pm\rangle$ states in fact contain an admixture of intermediate states
$|N-1,1\rangle,\ldots,|1,N-1\rangle$ such that their overlap is $\langle +
| - \rangle =\epsilon$. For $\epsilon\gg 1$ they shrink to

\begin{equation}\label{dirty}
(a^{\dagger} \pm b^{\dagger})^N \; |0,0\rangle \;\;.
\end{equation}
which does not correspond to a macroscopic superposition. From now on we 
assume the pure case, $\epsilon\ll 1$.

The proposal of Gordon and Savage \cite{gs} for obtaining these superposition
states is similar to a typical experiment in quantum optics with non-linear
systems. In the BEC case, the non-linearity is provided by the collisional
interactions and the Josephson coupling. The idea is to prepare an initial
state with all the atoms in the same internal state and then turn on the
Josephson coupling for some amount of time adequately chosen. After the
Josephson coupling is turned off, the quantum state of the system has been
modified, and a Schr\"odinger cat state has been formed. Other proposals
for obtaining basically the same kind of superposition use
a continuous quantum measurement process \cite{r2}.

\section{The thermal cloud}

As we have already mentioned, neither of the proposals for generating 
quantum superpositions in BEC addresses the crucial question of decoherence.
In fact, the condensate is an open quantum system which is in contact with
an environment of non-condensed thermal particles. This interaction
may be responsible for the loss of coherence between the components of the
quantum superposition states of the previous section. If the decoherence 
time is very small, then the existence of these states in BEC would be merely
of academic interest, since the expectations of observing them in the lab
would vanish. Therefore it is important to understand how the
thermal cloud affects the longevity of the BEC cats.

The Hamiltonian for the dilute environment formed by non-condensed particles
is

\begin{equation}\label{HEab}
H_{\rm E}=\sum_s\left[
\epsilon_s(a_s^{\dagger}a_s+b_s^{\dagger}b_s)
-\lambda(a_s^{\dagger}b_s+b_s^{\dagger}a_s)
\right]\;.
\end{equation}
$\epsilon_s$ is the single particle energy of level $s$. After the 
transformation

\begin{equation}
S_s=\frac{a_s+b_s}{\sqrt{2}} \;\;,\;
O_s=\frac{a_s-b_s}{\sqrt{2}} 
\end{equation}
$H_{\rm E}$ adopts a diagonal form

\begin{equation}\label{HEAB}
H_{\rm E}=\sum_s\left[
(\epsilon_s-\lambda)S_s^{\dagger}S_s+
(\epsilon_s+\lambda)O_s^{\dagger}O_s
\right]\;.
\end{equation}

In Figure 1 we make a schematic plot of the energy levels of the non-condensed
system before ($\lambda =0$) and after ($\lambda \neq 0$) applying the
Josephson coupling. For simplicity we consider an isotropic harmonic trap
with a dip in its center, where the condensate particles will be localized
(we shall return to this point later). We see that when the laser coupling
is applied, the states $S_s$ (which are symmetric under the exchange 
$a \leftrightarrow b$) and the states $O_s$ (which are antisymmetric under
the same exchange) feel the same, but shifted, trapping potentials. The
two ladders of energy eigenstates of the environment are separated by a gap
equal to $2 \lambda$. Each of the levels is occupied according to the thermal
Bose distribution. When the gap is much bigger that the typical thermal energy
$2 \lambda \gg k_{\rm B} T$, then the antisymmetric states are unoccupied.
We will see in the next section that this is a prerequisite for reducing
decoherence.

\begin{figure}
\centering \leavevmode
\epsfxsize=7cm
\epsfbox{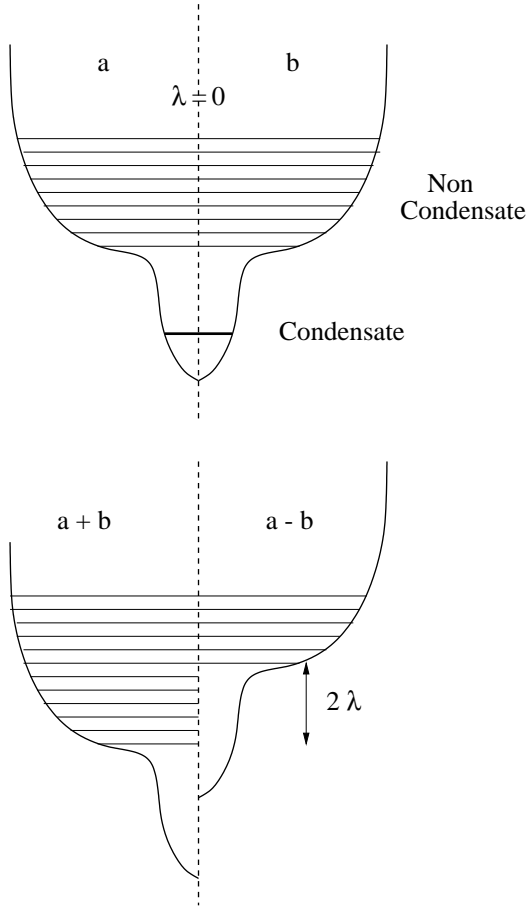}
\caption{Environmental single-particle energy levels before and after turning
on the Josephson-like coupling.}
\end{figure}

\section{Condensate-Non condensate interactions}

Condensed atoms interact with non-condensed atoms via two-body collisions.
The exact interaction hamiltonian $V$ can be easily obtained from the
total (Gross-Pitaevskii like) hamiltonian

\begin{equation}\label{H}
H=\int d^3x \left[ 
v(\phi_A^{\dagger}\phi_A)(\phi_B^{\dagger}\phi_B)+ 
\left\{
\frac{u}{2} \phi_A^{\dagger}\phi_A^{\dagger} \phi_A \phi_A +
\nabla\phi_A^{\dagger}\nabla\phi_A+
U(r)\phi_A^{\dagger}\phi_A-
\lambda \phi_A^{\dagger}\phi_B \right\}+                
\left\{ A \leftrightarrow B \right\} \right] ,
\end{equation}
after splitting the quantum fields $\phi_A$ and $\phi_B$ into condensate
and bath parts, i.e.

\begin{eqnarray}\label{phiAB}
\phi_{A}(\vec{x})\;=
\;a\;g(\vec{x})\;+\;
\sum_{s}\;a_{s}\;u_{s}(\vec{x})\;, \nonumber\\
\phi_{B}(\vec{x})\;=
\;b\;g(\vec{x})\;+\;
\sum_{s}\;b_{s}\;u_{s}(\vec{x}) ,
\end{eqnarray}
where $g(\vec{x})$ is the wave function of the condensate localized in the
dip.
After introducing the symmetric and antisymmetric operators $S_s$ and $O_s$
for the thermal bath, the interaction hamiltonian can be splitted into two
terms, $V=V_{\rm S}+ V_{\rm O}$, which are respectively symmetric and 
antisymmetric under 
the interchange $a \leftrightarrow b$ of system operators. For the
sake of conciseness, we shall refrain from writing down the whole expressions
for these two interaction hamiltonians. Suffice it to say that each of them
has two possible energy-conserving interaction verteces, both proportional
to the interaction couplings $u$ or $v$: 

\begin{itemize}

\item There are inelastic two-body collisions
that do not conserve the number of condensed particles (condensate feeding
and depletion processes). There are in turn two possible diagrams for these 
processes, and they are shown in Figure 2. The first one is proportional to the
occupation number of initial non-condensed states; since each occupation number
is proportional to fugacity $z\equiv e^{\beta \mu}$ (where $\beta$ is the 
inverse temperature of the thermal bath and $\mu$ its chemical potential), then
the diagram is $O(z^2)$. The second one is proportional to the occupation
number of the initial non-condensed state (that is, proportional to $z$), and
energy conservation brings about another fugacity factor, so the final result
is that this diagram is also $O(z^2)$ \cite{anglin,wallsroul}.

\begin{figure}
\centering \leavevmode
\epsfxsize=14cm
\epsfbox{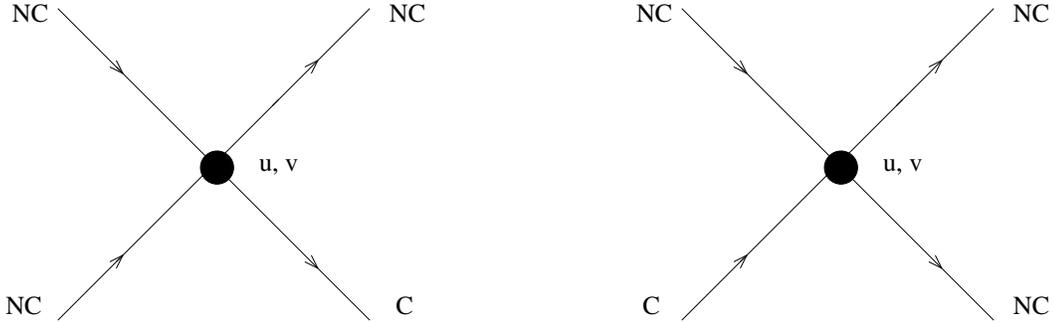}
\caption{Inelastic two-body collisions between condensate and non-condensate
particles}
\end{figure}

\item There are elastic two-body collisions that conserve the number
of condensed particles, and are shown in Figure 3. These diagrams are 
proportional to fugacity, $O(z)$ \cite{anglin,wallsroul}.

\begin{figure}
\centering \leavevmode
\epsfxsize=7cm
\epsfbox{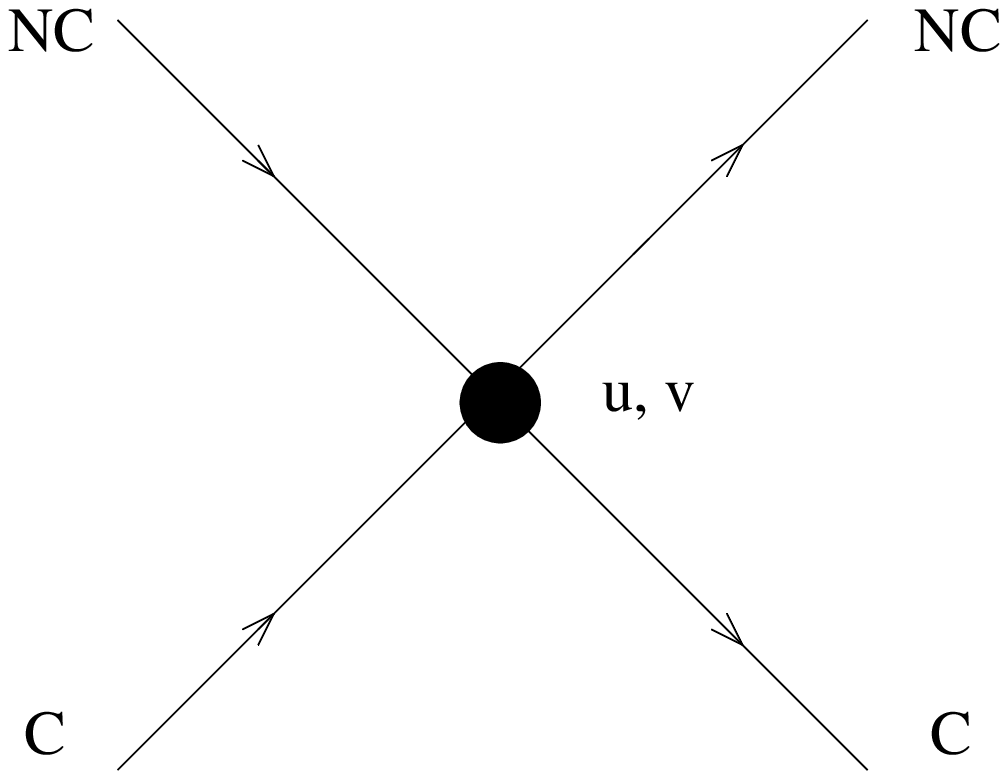}
\caption{Elsatic two-body collisions between condensate and non-condensate
particles}
\end{figure}

\end{itemize}

When fugacity is small (which occurs when the gap between the condensate
and non-condensate single-particle levels is bigger that the thermal energy
$k_{\rm B} T$), then the elastic processes dominate over the inelastic ones.
The elastic collisions are also the most relevant
ones in the computation of the decoherence rate, since we shall see that their
contribution scales as $N^2$ (with $N$ the total number of particles in the 
condensate), whereas the inelastic contribution scales as $N$. For these 
reasons, from now on we shall concentrate only on the elastic processes.

\section{Decoherence-free pointer subspace in BEC}

As we have mentioned in Section III, when $2 \lambda \gg k_{\rm B} T$ the
antisymmetric environmental states are nearly empty. Only the symmetric states
are occupied. Since these latter states cannot distinguish between $A$ and
$B$, all collisions that involve symmetric non-condensed states will not
destroy the quantum phase coherence between the Schr\"odinger cat's 
components. We refer to this condition as the perfect symmetrization limit.
In this case, the states 
$|\pm\rangle \equiv (|N,0\rangle\pm|0,N\rangle )/\sqrt{2}$ span a 
decoherence-free pointer subspace (also known in the literature
as decoherence-free subspace or, for short, DFS) of the Hilbert space, since 
they have 
degenerate eigenvalues of the interaction hamiltonian $V$ 
\cite{pointer,paolo,lidar}. 
Any state of that subspace can be written as
$\alpha |N,0\rangle + \beta |0,N\rangle$, with $\alpha$ and $\beta$ complex
numbers. If ${\cal{P}}_{[\alpha|N,0\rangle+\beta|0,N\rangle]}$ denotes a 
projector onto that subspace, then 

\begin{equation}
[V, {\cal{P}}_{[\alpha|N,0\rangle+\beta|0,N\rangle]}] = 0, 
\label{conmut}
\end{equation}
which means that any quantum superposition
$\alpha |N,0\rangle + \beta |0,N \rangle$ in the subspace 
is an eigenstate of the
interaction hamiltonian (a perfect pointer state), 
and as such will retain its phase
coherence and last forever. The
interaction Hamiltonian between the condensate and the thermal cloud is a
sum of products of condensate operators and environmental operators. Only
terms with symmetric environmental operators are relevant because
the antisymmetric states are empty.  
The total Hamiltonian is symmetric with respect to $A \leftrightarrow B$
so, to preserve this symmetry, the relevant terms with symmetric
environmental operators also contain symmetric condensate operators. The
argument simplifies a lot for small fugacity where there is only one
leading term with the $N_A+N_B$ condensate operator. The states
$|\pm\rangle$ are its eigenstates with the same eigenvalue $N$. They are
also (almost) degenerate eigenstates of the condensate Hamiltonian build
out of $N_{A,B}$. The coherent transitions $A\leftrightarrow B$ break this
degeneracy of $|\pm\rangle$ but the difference of their eigenfrequencies
is negligible as compared to the usual condensate lifetime of $\sim 10$s.
In the next-to-leading order in fugacity there are symmetric interaction
terms which change the number of condensed atoms. These terms drive the
$|N,0\rangle$ and $|0,N\rangle$ states into slightly ``squeezed-like'' states
$|S,0\rangle$ and $|0,S\rangle$ respectively \cite{moon}. 
There are also terms which exchange $A$ with
$B$. They give each state a small admixture of the opposite component. 
Superpositions of these are still decoherence-free pointer subspaces - 
there are no relevant antisymmetric operators to
destroy their quantum coherence.

\section{Decoherence comes into the scene}

When the antisymmetric environmental states begin to be occupied, 
then the commutation relation (\ref{conmut}) is only approximate
and states within the subspace will decohere. Hence we must face the 
problem of calculating decoherence rates due to interactions with the 
environment. 

We shall first consider the case of $\lambda=0$, which corresponds to no
Josephson-coupling being applied, and for which the two states $|\pm \rangle$
are exactly degenerate with respect to the condensate hamiltonian. In this
case there is no difference between symmetric and antisymmetric environmental
states. After
some long but straightforward calculation which involves computing the master
equation for the reduced density matrix of the condensed particles (see
\cite{us} for details), we find a lower bound for the decoherence rate 

\begin{equation}
t_{\rm dec}^{-1} > 
16 \pi^3\; \left( 4\pi a^2 \frac{N_{\rm E}}{V} v_T  \right) \; N^2  ,
\label{dr}
\end{equation}
where $N$ is a number of condensed atoms, $v_T=\sqrt{2k_{\rm B}T/m}$ is a
thermal velocity in the noncondensed thermal cloud, $a$ is a scattering
length, $V$ is a volume of the trap, and $N_{\rm E}$ 
is a number of atoms in the thermal cloud,

\begin{equation}\label{NE}
N_{\rm E}
\;\approx\;
e^{\frac{\mu}{k_{\rm B} T}} \;
\left( \frac{k_{\rm B}T}{\hbar\omega} \right)^3 .
\end{equation}
This is a lower estimate since we have only considered terms to leading
order in fugacity $O(z)$ and to leading order in condensate size $O(N^2)$.
Next-to-leading order terms are $O(z^2)$ and $O(N)$, in agreement with Refs.
\cite{anglin,wallsroul}, so they were neglected here.
Even without going into details of our derivation it is easy to understand 
where a formula like Eq.(\ref{dr}) comes from. $N^2$ is the main factor which 
makes the decoherence rate large. It comes from the master equation of the 
Bloch-Lindblad form
$\dot\rho\sim -[N_{\rm A}-N_{\rm B},[N_{\rm A}-N_{\rm B},\rho]]$, with
$\rm A$ and $\rm B$ the two internal states of the atoms. $N^2$ is the
distance squared between macroscopically different components of the
superposition $(|N,0\rangle + |0,N\rangle)/\sqrt{2}$ - 
the common wisdom reason why
macroscopic objects are classical \cite{zurek}. 
The factor in brackets in Eq.(\ref{dr}) is a scattering
rate of a condensate atom on noncondensate atoms - the very process by
which the thermal cloud environment learns the quantum state of the
system.

Let us estimate the decoherence rate for a set of typical parameters:
$T=1\mu$K, $\omega=50$Hz, and $a=3\dots 5$nm. The thermal velocity is
$v_{\rm T} \approx 10^{-2}$m/s. The volume of the trap can be approximated
by $V=4\pi a_{\rm return}^3/3$, where $a_{\rm return}=\sqrt{2 k_{\rm B}T/m
\omega^2}$ is a return point in a harmonic trap at the energy of $k_{\rm B}T$. 
We estimate the decoherence time as $t_{\rm dec} \approx 10^{5} 
{\rm sec} / ( N_{\rm E} N^2)$. For $N_{\rm E}=10^0\dots 10^4$ and $N=10\dots
10^7$ it can range from $1000$s down to $10^{-13}$s. For $N=10$ our
(over-)estimate for $t_{\rm dec}$ is large. However, already for $N=1000$
and $N_{\rm E}=10$ (which are still within the limits of validity of 
the two-mode approximation)
$t_{\rm dec}$ shrinks down to milliseconds. Given that our $t_{\rm dec}$
is an upper estimate and that big condensates are more interesting as
Schr\"odinger cats, it is clear that for the sake of cat's
longevity, one must go beyond the standard harmonic trap setting.

\section{Trap engineering}

From Eqs.(\ref{dr},\ref{NE}) it is obvious that the decoherence rate
depends a lot on temperature and on chemical potential. The two factors
strongly influence both $N_{\rm E}$ and $v_{\rm T}$. Both can be improved
by the following scenario, which is a combination of present day
experimental techniques. In the experiment of Ref.\cite{dip} a narrow
optical dip was superposed at the bottom of a wide magnetic trap. The
parameters of the dip were tuned so that it had just one bound state. The
gap between this single condensate mode and the first excited state was
$1.5\mu$K, which at $T=1\mu$K gives a fugacity of $z=\exp(-1.5)$. 
We need the gap so that we can use the single mode approximation. At low
temperatures, the gap results in a small fugacity, which is convenient 
for calculations.
We propose to prepare a condensate inside a similar combination of a wide
magnetic and a narrow optical trap (or more generally: a wide well plus a
narrow dip with a single bound mode) and then to open the magnetic trap
and let the noncondensed atoms disperse. The aim is to get rid of the
thermal cloud as much as possible. A similar technique was used in the
experiment of Ref.\cite{optical} (see Figure 4).

\begin{figure}
\centering \leavevmode
\epsfxsize=16cm
\epsfbox{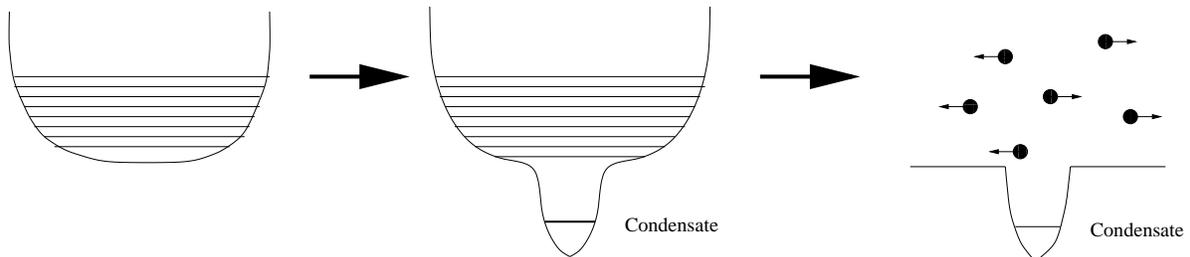}
\caption{First step in engineering the trap: an optical dip is superimposed
on a big magnetic trap, which is later opened so that most of the thermal
atoms disperse away.}
\end{figure}

Let us estimate the ultimate limit for the efficiency of this technique.  
At the typical initial temperature of $1\mu$K the thermal velocity of
atoms is $10^{-2}$m/s. An atom with this velocity can cross a $1\mu$m dip
in $10^{-4}$s. If we wait for, say, $1$s after opening the wide trap, then
all atoms with velocities above $10^{-6}$m/s will disperse away from the
dip.  A thermal velocity of $10^{-6}$m/s corresponds to the temperature of
$10^{-8}\mu$K. As the factor $N_{\rm E} 
v_T \sim T^{7/2}$ in Eq.(\ref{dr}), then
already $1$s after opening the wide trap the decoherence rate 
due to non-condensed atoms is reduced
by a factor of $10^{-28}$!

It is not realistic to expect such a ``cosmological" reduction factor.
The ``dip" which is left after the wide harmonic trap is gone could be, for
example, a superposition of an ideal dip plus a wide shallow well (which
was a negligible perturbation in presence of the wide harmonic trap).  
The well would have a band of width $\Delta E$ of bound states which would
not disperse but preserve their occupation numbers from before the opening
of the wide trap. They would stay in contact with the condensate and continue
to ``monitor'' its quantum state. 
Even if such a truncated environment happens
to be already relatively harmless, there are means to do better than that.

Further reduction of the decoherence rate can be achieved by 
``symmetrization" of the environmental states. To this end we propose
to turn on the Josephson-like coupling $\lambda$ in order to induce
coherent transitions from states A and B. This has already been achieved
experimentaly \cite{lambda}; in their case $\lambda \approx 1 {\rm kHz}$.
When $\Delta E \ll 2 \lambda$ we are in the perfect symmetrization limit, and
hence no decoherence takes place. Indeed, the symmetric and
antisymmetric $\Delta E$-bands of states can be visualized as two ladders
shifted with respect to each other by $2 \lambda$. In other words, the two
sets of states feel the same, but shifted, trapping potentials. When
$\Delta E \ll 2 \lambda$, then the antisymmetric $O_s$'s are nearly 
empty since they can
evaporate into symmetric states and then leave the trap . 
The symmetric
$S_s$'s cannot distinguish between $A$ and $B$ so they do not destroy the
quantum coherence between the Schr\"odinger cat's components. After
symmetrization, $N_{\rm E}$ in Eq.(\ref{dr}) has to be replaced by the
final number of atoms in the antisymmetric states only;

\begin{equation}\label{NEO}
N_{\rm E}^{\rm O}
\;\approx\;
\left\{
\begin{array}{ll}
n_{\lambda} \exp( (\mu-\lambda)/ k_{\rm B} T ),  
& \mbox{for $2 \lambda < \Delta E$} \\
0,  & \mbox{for $2 \lambda > \Delta E$}
\end{array}
\right.
\end{equation}
Here $T$ is the temperature before opening the wide trap. $n_{\lambda}$ is
the number of antisymmetric bound states which remain within the $\Delta
E$-band of symmetric states. Atoms in these antisymmetric bound states
cannot disperse away. For $2\lambda > \Delta E$ this number $n_{\lambda}$
is zero and there is no decoherence from the thermal cloud.

In Figure 5 we make a schematic graph of the steps we propose to engineer the
trap in order to reduce decoherence due to the thermal cloud.

\begin{figure}
\centering \leavevmode
\epsfxsize=14cm
\epsfbox{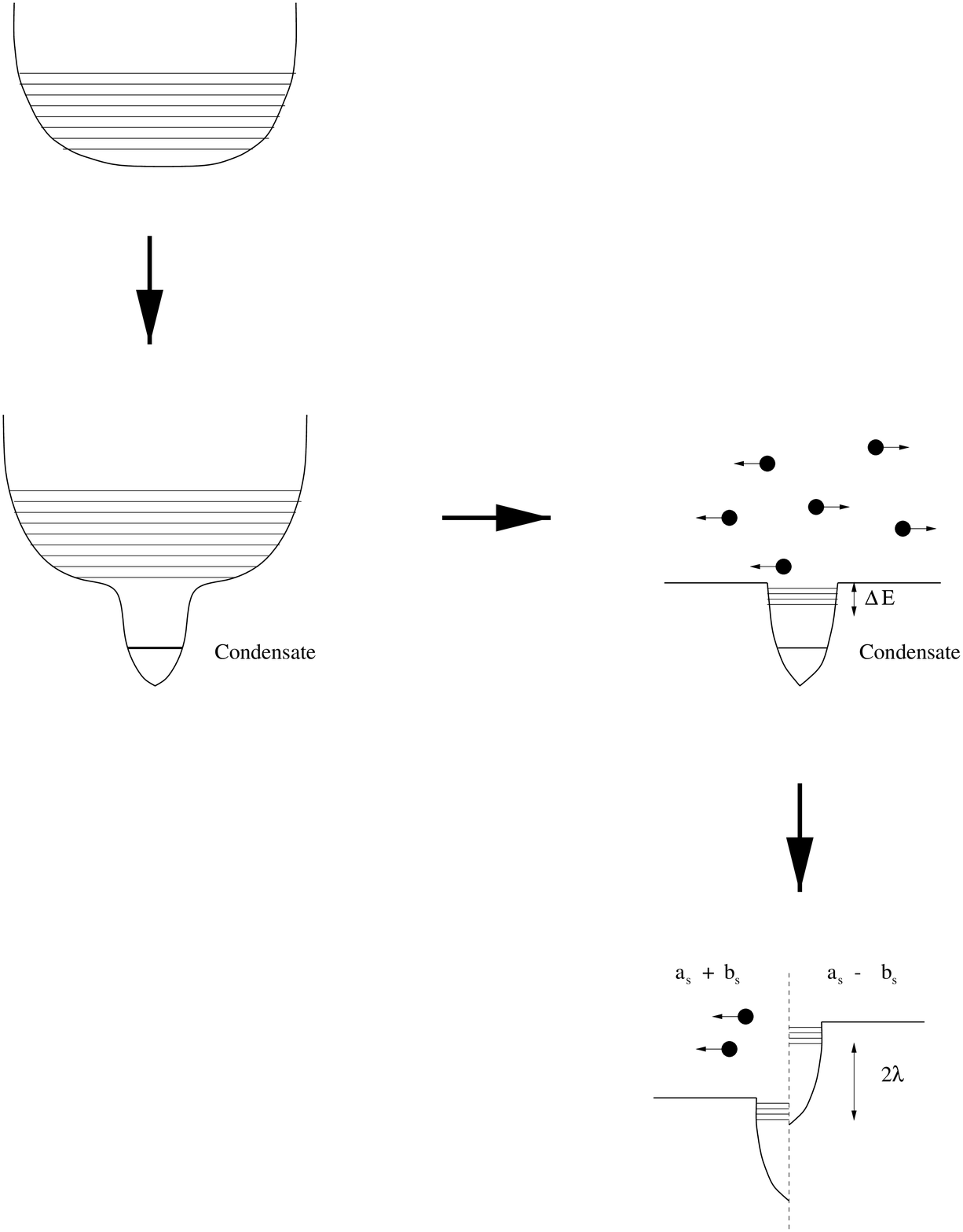}
\caption{Second step in engineering the trap: After the magnetic trap is 
opened, there is still a band $\Delta E$ of thermal states in the ``mouth''
of the dip. Turning on the Josephson coupling $\lambda$ shifts the energy
levels of symmetric and antisymmetric thermal states, and the latter become
empty.}
\end{figure}

\section{Other sources of decoherence}

Interaction with the thermal atoms is not the only source of decoherence
for the quantum superposition state of the condensate. Among other possible
sources we can mention:

\begin{itemize}

\item Ambient magnetic fields: the condensed atoms have magnetic moments. If 
the magnetic moments of A and B were different the magnetic field would 
distinguish between them and would introduce an unknown phase into the
quantum superposition, thus rendering its underlying coherent nature 
undetectable. Fortunately the much used 
$|F,m_F \rangle=|2,1\rangle,|1,-1\rangle$ states of 
$^{87}$Rb have the same magnetic moments. For them the magnetic field
is a ``symmetric" environment.

\item Different scattering lengths: there is a typical $\approx 1 \%$ 
difference
between the $A-A$ and $B-B$ scattering lengths. The condensate Hamiltonian is 
not perfectly symmetric under $A\leftrightarrow B$. Even for a perfectly 
symmetrized environment,
symmetric environmental operators couple to not fully symmetric condensate
operators.  This means that for the $1 \%$ difference of scattering lengths
symmetrization can improve decoherence time by at most two orders of
magnitude as compared to the unsymmetrized environment.

\item Three-body losses: due to collisions involving three particles,
the condensate is not stable but loses atoms via recombination into
molecular states. Hence a condensate has a finite lifetime, of the 
order of $10\dots 20$s. The atoms which escape from the
condensate carry information about its quantum state. They destroy its
quantum coherence. In the experiment of Ref.\cite{dip}
the measured loss rate per atom was $4/$s for $N=10^7$ or around $1$ atom
per $10^{-7}$s. The last rate scales like $N^3$ so already for $N=10^4$
just one atom is lost per second; decoherence time is $1$s. One possibility
for increasing this decoherence time is
to increase slightly the dip radius. The loss rate scales
like density squared so an increase in the dip width by a factor of $2$
reduces the loss rate by a factor of $2^6=64$.

\end{itemize}

\section{Discussion}

The aim of this paper was to discuss the ``longevity'' of Schr\"odinger
cats in BEC's. We have shown that while in the standard traps decoherence
rates are significant enough to prevent long-lived macroscopic superpositions
of internal states of the condensate, the strategy of trap engineering and
symmetrization of the
environment will be able to deal with that issue. 

What remains to be considered is how  one can generate such macroscopic quantum
superposition, and how one can detect it. The issue of generation was
already touched upon in Refs. \cite{lz,gs}. We have little to add to this.
However, in the context of the Gordon and Savage proposal, it is fairly
clear that the time needed to generate the cat state would have to be short
compared to the decoherence time. If our estimates of Eq.(\ref{dr}) are 
correct, symmetrization procedure appears necessary for the success of such 
schemes.

Detection of Schr\"odinger cat states is perhaps a more challenging subject.
In principle, states of the form $( |N,0\rangle + |0,N\rangle) /\sqrt{2}$
have a character of GHZ states, and one could envision performing 
measurements analogous to those suggested in \cite{molmer} and
carried out in \cite{four}, where a 4-atom entangled state was studied.
However, this sort of parity-check strategy, appropriate for $N\leq 10$,
is likely to fail when $N$ is larger, or when (as would be the case for
the ``quasi-squeezed'' states anticipated here \cite{moon}) $N$ is not
even well defined. 

A strong circumstantial evidence can be nevertheless obtained from two
measurements. The first one would consist of a preparation of the cat state,
and of a measurement of the internal states of the atoms. It is expected that
in each instance all (to within the experimental error) would turn up to be
in either A or B states. However, averaged over many runs, the number of either
of these two alternatives would be approximately equal. Decoherence in which
the environment also ``monitors'' the internal state of the atoms in the
A versus B basis would not influence this prediction. We need to check
separately whether the cat state was indeed coherent. To do this, one could
evolve the system ``backwards''. However, this is not really necessary. For, 
as Gordon and Savage point out, 
when, in their scheme,
we let the system evolve unitarily
for more or
less twice
the time needed for the generation of the cat state, it will approximately
return to the initial configuration. Thus, we can acquire strong evidence
of the coherence of the cat provided that this 
unitary
return to the initial
configuration can be experimentally confirmed.

These are admittedly rather vague ideas, which serve more as a ``proof of
principle'' rather than as a blueprint for an experiment. Nevertheless,
they may, we hope, encourage more concrete investigation of 
such issues with a specific experiment in mind.

\section{Acknowledgements} 

We are indebted to E.Cornell, R.Onofrio, E.Timmermans, and especially to
J.Anglin for very useful comments. This research was supported in part by
NSA.

\end{document}